\documentstyle[11pt,epsf]{article}
\newtheorem{theorem}{Theorem}

\newcommand {\dfn} {\stackrel{\Delta} {=}}
\newcommand {\exe} {\stackrel{\cdot} {=}}
\newcommand {\lexe} {\stackrel{\cdot} {\le}}

\newcommand {\reals} {{\rm I\!R}}

\newcommand {\bx} {\mbox{\boldmath $x$}}
\newcommand {\by} {\mbox{\boldmath $y$}}

\newcommand {\bE} {\mbox{\boldmath $E$}}

\newcommand {\hP} {\hat{P}}

\newcommand {\tX} {\tilde{X}}
\newcommand {\tW} {\tilde{W}}

\newcommand {\bX} {\mbox{\boldmath $X$}}

\newcommand{\calC}{{\cal C}}

\newcommand{\calE}{{\cal E}}

\newcommand{\calI}{{\cal I}}

\newcommand{\calT}{{\cal T}}

\newcommand{\calX}{{\cal X}}
\newcommand{\calY}{{\cal Y}}

\begin{document}
\thispagestyle{empty}
\title{A Lagrange--Dual Lower Bound to the Error Exponent\\
Function of the Typical
Random Code\thanks{This research was supported by the Israel Science Foundation (ISF),
grant no.\ 137/18.}}
\author{Neri Merhav}
\date{}
\maketitle

\begin{center}
The Andrew \& Erna Viterbi Faculty of Electrical Engineering\\
Technion - Israel Institute of Technology \\
Technion City, Haifa 32000, ISRAEL \\
E--mail: {\tt merhav@ee.technion.ac.il}\\
\end{center}
\vspace{1.5\baselineskip}
\setlength{\baselineskip}{1.5\baselineskip}

\begin{abstract}
A Lagrange--dual (Gallager--style) lower bound is derived for the error
exponent function of the typical random code (TRC) pertaining to the i.i.d.\ random coding
ensemble and mismatched stochastic likelihood decoding.
While the original expression, derived from the method of types (the
Csisz\'ar--style expression) involves minimization over probability
distributions defined on the channel input--output alphabets, the new Lagrange--dual
formula involves optimization of five parameters, independently of the
alphabet sizes. For both stochastic and deterministic mismatched decoding
(including maximum
likelihood decoding as a special case), 
we provide a rather comprehensive discussion on the insight behind the various
ingredients of this formula and describe how its
behavior varies as the coding rate exhausts the relevant range. Among other
things, it is demonstrated that this expression simultaneously generalizes both the expurgated
error exponent function (at zero rate) and the classical random coding
exponent function at
high rates, where it also meets the sphere--packing bound.\\

\end{abstract}

\newpage
\section{Introduction}

In view of the articles by Barg and Forney \cite{BF02}, Nazari
\cite{Nazari11} and Nazari {\it et al.} \cite{NAP14}, in a more recent paper
\cite{trc}, the exact error exponent function of the typical random code (TRC)
for a given discrete memoryless channel (DMC) was derived analytically for
the ensemble of fixed--composition codes. The
error exponent of the TRC
was defined as the limit 
of the negative normalized {\it expectation of the logarithm} of the error
probability, which is different from the classical random coding exponent, defined as the
negative normalized {\it logarithm of the expectation} of the error
probability, where both expectations are with respect to (w.r.t.) 
the randomness of the code. This study of TRC error exponents
was motivated by various considerations. The first is that due to
Jensen's inequality, it is always greater than or equal to the random coding error exponent, and
it is therefore a more optimistic performance metric than the classical random
coding exponent, especially
at a certain range of low rates. The second consideration is that whenever a certain measure concentration
property holds, it is a more relevant figure of merit, because
the code is normally assumed to be randomly selected only once,
and then it is used repeatedly. Last but not least, it is coherent with
the notion of
random--like codes \cite{Battail95},
which are considered very good codes.

In \cite{trc},
an exact single--letter expression was derived for the error exponent function
of the TRC assuming a general finite alphabet, discrete memoryless channel (DMC) w.r.t.\ the
ensemble of fixed composition codes and a family of stochastic decoders,
referred to as generalized likelihood decoders (GLDs), which includes
many relevant deterministic decoders (like 
the maximum likelihood decoder) as special cases. Among other things, it was shown
in \cite{trc} (similarly as in \cite{BF02} and \cite{Nazari11}),
that the error exponent function of the TRC has the following properties: (i) it agrees
with the expurgated exponent at
rate zero, (ii) it is smaller than the expurgated exponent, but
larger than the random coding exponent at a certain range low rates, and (iii)
it coincides with 
the random coding exponent above a certain rate. Other, more recent follow--up papers
related to TRC exponents, include time--varying trellis codes \cite{trellis-trc}, codes
for colored Gaussian channel \cite{trcg}, joint source--channel coding
\cite{trcjsc}, and large deviations about the TRC exponent
\cite{largedeviations}.

The error exponent formula of \cite[Theorem 1]{trc} was derived using the method of
types, and as such, it was presented in terms of several (nested) optimizations
of a certain information--theoretic expression over some joint probability distributions
and conditional distributions whose support depends on the channel input and
output alphabets. We henceforth refer to this expression as the
{\it Csisz\'ar--style} formula. The total number of parameters w.r.t.\ which this
expression should be optimized is $|\calX|^2\cdot|\calY|+(|\calX|-1)\cdot|\calY|-1$,
where $|\calX|$ and $|\calY|$ are the cardinalities of the input alphabet and
the output alphabet, respectively. Even in the simplest case of a
binary--input/binary--output channel ($|\calX|=|\calY|=2$),
this number is already as large as nine, and it grows, of course, extremely
rapidly as the alphabet sizes grow. Moreover, out of this number of
parameters, $|\calX|^2\cdot|\calY|-1$ are associated with minimization and 
the remaining $(|\calX|-1)\cdot|\calY|$ parameters undergo maximization. We make this distinction because if one is
interested merely in guaranteed performance, namely, a valid lower bound to the
error exponent, there is complete freedom in choosing the latter parameters in an
arbitrary manner (rather than maximizing over them), but there is still a necessity to find the global minimum over the
former $|\calX|^2|\calY|-1$ parameters, which is still computationally very
demanding even for moderate alphabet sizes. 

These facts motivate us to derive the Lagrange--dual of the above--mentioned Csisz\'ar--style
formula, a.k.a.\ the {\it Gallager--style} formula, which is technically, the main result
of this paper. For the sake of simplicity, we derive the
Lagrange--dual expression for the ensemble of codes drawn from an i.i.d.\
distribution, in contrast to the ensemble of fixed composition codes, used in
\cite{trc}. Although the i.i.d.\ ensemble cannot be better than the fixed
composition ensemble \cite[Sect.\ IV.D]{trc}, we opt to adopt the former for several reasons.
\begin{enumerate}
\item The derivation of the Lagrange--dual expression is considerably easier
and simpler for the i.i.d.\
ensemble than for the fixed--composition ensemble, because it is free of
the constraints associated with the
fixed composition assumption. While the Lagrange--dual form
can be derived for the fixed composition too, the
cost of eliminating those fixed--composition constraints is in having many additional parameters for
optimization.
\item The i.i.d.\ ensemble is very important on its own right. It has been investigated in much of
the earlier, classical work on error exponents \cite{Gallager68}. In
particular, it was studied \cite{BF02} for TRC
exponents (among other things), albeit
only for the special case of the binary symmetric channel.
\item At least at zero--rate and above the critical rate, there is provably no
loss of optimality, at least when the random coding distribution is chosen
optimally.
\end{enumerate}

While the resulting Lagrange--dual expression is still not
trivial, it is nevertheless computationally preferable by far relative to the
Csisz\'ar--style expression,
as it is associated with optimizations over five parameters, independently of
the alphabet sizes. Four of these optimizations are
maximizations, and only one is a minimization. Thus, as
described above, for a valid lower bound on the TRC exponent,
we have full freedom in choosing the former four parameters,
and for only one parameter, it is necessary to
conduct a minimization. 

But the benefit of the Lagrange--dual, Gallager--style lower bound to the TRC
exponent is not limited
to computational aspects alone. 
We also provide a rather comprehensive discussion on the
insight behind the various
ingredients of this formula and describe how its
behavior varies as the coding rate exhausts the relevant range. Among other
things, it is demonstrated that this expression simultaneously generalizes
both the expurgated
error exponent function (at zero rate) and the classical random coding
exponent function, at
high rates, where it also meets the sphere--packing bound.

The outline of the remaining part of this paper is as follows. In Section
\ref{nab}, we establish notation conventions, define the setup, and provide
the necessary background from \cite{trc}. In Section \ref{mrd}, we
present the main result -- the Lagrange--dual lower bound to the TRC exponent, 
and discuss it. Finally, in Section \ref{proof}, we
prove the Lagrange--dual formula.

\section{Notation and Background}
\label{nab}
\subsection{Notation}

Throughout the paper, random variables will be denoted by capital
letters, specific values they may take will be denoted by the
corresponding lower case letters, and their alphabets
will be denoted by calligraphic letters. Random
vectors and their realizations will be denoted,
respectively, by capital letters and the corresponding lower case letters,
both in the bold face font. Their alphabets will be superscripted by their
dimensions. For example, the random vector $\bX=(X_1,\ldots,X_n)$, ($n$ --
positive integer) may take a specific vector value $\bx=(x_1,\ldots,x_n)$
in $\calX^n$, the $n$--th order Cartesian power of $\calX$, which is
the alphabet of each component of this vector.
Sources and channels will be denoted by the letters $P$, $Q$ and $W$,
subscripted by the names of the relevant random variables/vectors and their
conditionings, if applicable, following the standard notation conventions,
e.g., $Q_X$, $P_{Y|X}$, and so on. When there is no room for ambiguity, these
subscripts will be omitted.
The expectation
operator with respect to (w.r.t.) a probability distribution $Q$ will be
denoted by
$\bE_Q\{\cdot\}$. Again, the subscript will be omitted if the underlying
probability distribution is clear from the context.
For two positive sequences, $\{a_n\}$ and $\{b_n\}$, the notation $a_n\exe b_n$ will
stand for equality in the exponential scale, that is,
$\lim_{n\to\infty}\frac{1}{n}\log \frac{a_n}{b_n}=0$. Similarly,
$a_n\lexe b_n$ means that
$\limsup_{n\to\infty}\frac{1}{n}\log \frac{a_n}{b_n}\le 0$, and so on.
The indicator function
of an event $\calE$ will be denoted by $\calI\{E\}$. The notation $[x]_+$
will stand for $\max\{0,x\}$.

The empirical distribution of a sequence $\bx\in\calX^n$, which will be
denoted by $\hP_{\bx}$, is the vector of relative frequencies 
of each symbol $x\in\calX$ in $\bx$.
Similarly, the joint empirical distribution of a sequence pair,
$(\bx.\by)\in\calX^n\times\calY^n$, will be denoted by $\hP_{\bx\by}$.
The type class of a vector $\bx^n$ with empirical distribution $\hP_{\bx}=Q_X$
will be denoted by
$\calT(Q_X)$, and similarly the type class of a pair of vectors
$(\bx,\tilde{\bx})$ with joint empirical distribution $Q_{X\tX}$ will be
denoted by $\calT(Q_{X\tX})$.
For a generic distribution, $Q_{XY}$ (or $Q$, for short, when there is no risk of
ambiguity), we use the following notation for information measures:
$H_Q(X)$ -- for the entropy of $X$,
$H_Q(X,Y)$ -- for the joint entropy,
$H_Q(X|Y)$ -- for the conditional entropy of $X$ given $Y$,
$I_Q(X;Y)$ -- for the mutual information, and similar conventions
for other information measures and
for joint distributions of more than two random variables.
We will also use the customary notation for the weighted divergence,
\begin{equation}
D(Q_{Y|X}\|P_{Y|X}|Q_X)=\sum_{x\in\calX}Q_X(x)\sum_{y\in\calY}Q_{Y|X}(y|x)\log
\frac{Q_{Y|X}(y|x)}{P_{Y|X}(y|x)}.
\end{equation}

\subsection{Background}

Consider a DMC, $W=\{W(y|x),~x\in\calX,~y\in\calY\}$, where $\calX$ is a
finite input alphabet, $\calY$ is a finite output alphabet, and $W(y|x)$ is
the channel input--output single--letter transition probability from $x$ to $y$.
When  fed by a
vector $\bx=(x_1,x_2,\ldots,x_n)\in\calX^n$, the channel responds by producing
an output vector $\by=(y_1,y_2,\ldots,y_n)\in\calY^n$, according to
\begin{equation}
W(\by|\bx)=\prod_{i=1} W(y_i|x_i).
\end{equation}
Let $\calC_n=\{\bx_0,\bx_1,\ldots,\bx_{M-1}\}\subseteq\calX^n$, $M=e^{nR}$,
$R$ being the
coding rate in nats per channel use. When the transmitter wishes to convey
a message $m\in\{0,1,\ldots,M-1\}$, it feeds the channel with $\bx_m$. In
\cite{trc}, we
considered the ensemble of fixed composition codes, where each codeword is
selected independently at random under the uniform distribution across a
given type class of $n$--vectors, $\calT(Q_X)$.

As in \cite{gld} and \cite{trc}, we adopt an extended
version of the so called {\it likelihood decoder}
\cite{SMF15}, \cite{SCP14}, \cite{YAG13}, which is a stochastic decoder that
randomly selects the message estimate according to the posterior
probability distribution given $\by$. The generalized likelihood decoder (GLD)
randomly selects the decoded message $\hat{m}$ according to the
generalized posterior,
\begin{equation}
P(\hat{m}=m|\by)=\frac{\exp\{ng(\hP_{\bx_m\by})\}}{\sum_{m^\prime=0}^{M-1}
\exp\{ng(\hP_{\bx_{m^\prime}\by})\}},
\end{equation}
where  the function $g(\cdot)$, henceforth referred to as
the {\it decoding metric}, is a continuous function that maps joint probability
distributions over $\calX\times\calY$ to real numbers.
Thus, considering the function
$g(Q)=\beta\cdot\bE_Q\ln W(Y|X)$ (for a given $\beta > 0$), the choice $\beta=1$
corresponds to the ordinary posterior of $m$ given $\by$, and the limit $\beta\to\infty$ 
yields the deterministic maximum a--posteriori (MAP) decoder, which is also the maximum likelihood
(ML) decoder in this case. The choice $g(Q_{XY})=\beta\cdot\bE_Q\ln
\tilde{W}(Y|X)$, where $\tilde{W}$ is a possibly different channel,
corresponds to a family of stochastic mismatched decoders, which we will
adopt throughout this paper. Once again,
the limit $\beta\to\infty$ gives a deterministic decoder, in this case, a
mismatched decoder.
Other interesting choices of $g$ were discussed in \cite{gld}, \cite{trc} as
well as in other earlier works.

The probability of error, associated with a given code $\calC_n$ and the GLD,
is given by
\begin{equation}
\label{explicit-pe}
P_{\mbox{\tiny e}}(\calC_n)=\frac{1}{M}\sum_{m=0}^{M-1}\sum_{m^\prime\ne
m}\sum_{\by\in\calY^n}W(\by|\bx_m)\cdot
\frac{\exp\{ng(\hP_{\bx_{m^\prime \by}})\}}{\sum_{\tilde{m}=0}^{M-1}
\exp\{ng(\hP_{\bx_{\tilde{m}}\by})\}}.
\end{equation}
For the ensemble of rate--$R$ fixed composition codes of type $Q_X$,
we define the TRC error exponent, associated with the decoding metric $g$,
according to
\begin{equation}
\label{trc-def}
E_{\mbox{\tiny trc}}^g(R,Q_X)=\lim_{n\to\infty}\left[-\frac{\bE\ln
[P_{\mbox{\tiny e}}(\calC_n)]}{n}\right],
\end{equation}
where the expectations are w.r.t.\ the randomness of $\calC_n$.
Note that $E_{\mbox{\tiny trc}}^g(R,Q_X)$ is defined in terms of the expectation of the
logarithm of the error probability, as opposed to the definition of the
ordinary random coding exponent, which is in terms of the logarithm of the
expectation of the error probability.

For a given distribution, $Q_Y$, over the channel output alphabet, let
\begin{equation}
\label{alpha-def}
\alpha(R,Q_Y)\dfn\sup_{\{Q_{\tilde{X}|Y}:~I_Q(\tilde{X};Y)\le R,~
Q_{\tilde{X}}=Q_X\}}[g(Q_{\tilde{X}Y})-I_Q(\tilde{X};Y)]+R,
\end{equation}
and
\begin{eqnarray}
\label{Gamma-def}
\Gamma(Q_{XX^\prime},R)&\dfn&\inf_{Q_{Y|XX^\prime}}\{D(Q_{Y|X}\|W|Q_X)+I_Q(X^\prime;Y|X)+\nonumber\\
& &[\max\{g(Q_{XY}),\alpha(R,Q_Y)\}-g(Q_{X^\prime Y})]_+\}.
\end{eqnarray}
The main result of \cite{trc} is the following.
\begin{theorem} \cite[Theorem 1]{trc}
Consider the setting described above. Then,
\begin{equation}
\label{trcee}
E_{\mbox{\tiny trc}}^g(R,Q_X)=\inf_{\{Q_{X^\prime|X}:~I_Q(X;X^\prime)\le
2R,~Q_{X^\prime}=Q_X\}}\{\Gamma(Q_{XX^\prime},R)+
I_Q(X;X^\prime)-R\}.
\end{equation}
\end{theorem}
As can be seen, the calculation of $\alpha(R,Q_Y)$, which is associated with
maximization over $Q_{\tilde{X}|Y}$, involves $|\calX|-1$ free parameters for each
$y\in\calY$, thus a total of $(|\calX|-1)\cdot|\calY$ parameters. The
minimizations over
$Q_{Y|XX^\prime}$ and $Q_{XX^\prime}$, are equivalent to one
minimization over $Q_{XX^\prime Y}$, which has $|\calX|^2\cdot|\calY|-1$ free
parameters, as described in the Introduction.

\section{Main Result and Discussion}
\label{mrd}

\subsection{Main Result}

We consider the same setting as defined in Section \ref{nab}, except that the
fixed composition ensemble is replaced by the i.i.d.\ ensemble, where the $M$
codewords are drawn independently, and each one is drawn under the product
distribution 
\begin{equation}
\label{product}
P(\bx)=\prod_{i=1}^n P(x_i).
\end{equation}
The corresponding TRC exponent is defined as in (\ref{trc-def}), where the
expectation is now taken w.r.t.\ the i.i.d.\ ensemble defined by $P$, and
it will be denoted by $E_{\mbox{\tiny trc}}^g(R,P)$.

Our main result is the following lower bound to the TRC exponent.

\begin{theorem}
\label{thm2}
Consider the setting defined in Section \ref{nab}, but with the
fixed--composition ensemble being replaced by the i.i.d.\ ensemble defined by
$P$ and a GLD with the decoding metric $g(Q)=\beta\bE_Q\ln\tilde{W}(Y|X)$, for
a given $\beta > 0$. Then,
\begin{eqnarray}
E_{\mbox{\tiny trc}}^g(R,P)&\ge& 
\sup_{0\le \sigma\le \beta}\sup_{0\le \tau\le\beta-\sigma}\inf_{\lambda\ge 0}\sup_{\theta\ge 0}
\sup_{\zeta\ge 1+\theta}
\left[-\zeta\ln\left\{\sum_{x\in\calX}P(x)\left[\sum_{x^\prime\in\calX}
P(x^\prime)\times\right.\right.\right.\nonumber\\
& &\left.\left.\left.\left(\sum_{y\in\calY}W(y|x)\cdot\frac{\tilde{W}^{\sigma+\tau}(y|x^\prime)}
{\tilde{W}^{\sigma}(y|x)\left[\sum_{\tilde{x}}P(\tilde{x})
\tilde{W}^{1/\lambda}(y|\tilde{x})\right]^{\lambda
\tau}}\right)^{1/(1+\theta)}\right]^{(1+\theta)/\zeta}
\right\}-\right.\nonumber\\
& &\left.(\zeta+\theta-\lambda\tau)R\right].
\end{eqnarray}
\end{theorem}

\subsection{Discussion}

First, we note that
Theorem \ref{thm2} provides a lower bound to the TRC exponent, unlike Theorem
1 of \cite{trc} that claims the exact TRC exponent. The reason is that, in
contrast to \cite{trc}, here, for the i.i.d.\ ensemble, we have not proved a matching upper bound,
because our emphasis in this work is on a dual expression for the guaranteed
performance of the typical random code. Also, in following paragraphs of our discussion,
we will discuss several specific choices of the parameters $\sigma$, $\tau$,
$\theta$, and $\zeta$, rather than maximizing upon them, and so, the
resulting expression can only be claimed to be a lower bound anyway. Having said that, we will
see in the sequel, that at least in certain situations, the resulting
quantities will turn out to be tight, as they will meet well known converse
bounds. For the matched case, they will also be coherent with results derived in earlier works.

While the lower bound in Theorem \ref{thm2} seems to be considerably complicated, some useful insights can
nevertheless be
gained using a few observations.\footnote{It should be noted that there is a certain similarity to the
dual expression derived in \cite{trcjsc} for Slepian--Wolf binning, but there are also quite
a few differences.}
In particular, first observe that the inner--most sum over
$y$ can be rewritten slightly differently as follows:
\begin{equation}
\label{innersum}
\sum_y
W(y|x)\cdot\left[\frac{\tilde{W}(y|x^\prime)}{\tilde{W}(y|x)}\right]^{\sigma}\cdot
\left[\frac{\tilde{W}(y|x^\prime)}{\left\{\sum_{\tilde{x}}P(\tilde{x})
\tilde{W}^{1/\lambda}(y|\tilde{x})\right\}^\lambda}\right]^{\tau}.
\end{equation}
For simplicity, consider the limit $\beta\to\infty$, where the GLD becomes
the deterministic mismatched decoder $\hat{m}=\mbox{arg
max}_m\tilde{W}(\by|\bx_m)$, and then the minimizations over $\sigma$ and
$\tau$ both become over all positive reals.
Let us think of the error event as the {\it disjoint} union of the events
$$\left\{\tilde{W}(\by|\bx_{m^\prime})> \max\left\{\tilde{W}(\by|\bx_m),\max_{\tilde{m}\ne
(m,m^\prime)}
\tilde{W}(\by|\bx_{\tilde{m}})\right\}\right\},~~~~~~m^\prime=1,2,\ldots,m-1,m+1,\ldots,M$$
where $\bx_m$ is the correct codeword.
In the above expression of the TRC exponent, $x$ represents the 
correct codeword $\bx_m$, $x'$ stands for $\bx_{m^\prime}$, and 
$\tilde{x}$ designates the codeword $\bx_{\tilde{m}}$ with the highest 
score among all competing wrong codewords other than $\bx_{m^\prime}$. 
The summation over $y$ in (\ref{innersum}) can be thought of as a single--letter version of
the Chernoff bound for 
the probability of the above event, which can be rewritten as 
$$\left\{\tilde{W}(\by|\bx_{m^\prime})>\tilde{W}(\by|\bx_m)\right\}\bigcap 
\left\{\tilde{W}(\by|\bx_{m^\prime})> 
\max_{\tilde{m}\ne (m,m^\prime)} \tilde{W}(\by|\bx_{\tilde{m}})\right\}.$$
Low coding rates are characterized by the regime where pairwise error events
dominate the error probability. These involve merely the competition between $\bx_m$
and $\bx_{m^\prime}$, just like in the simple union bound. In this case, the
TRC exponent is achieved for $\tau=0$,
and eq.\ (\ref{innersum}) has the meaning of the expectation (w.r.t.\ $W(y|x)$)
of $[\tilde{W}(y|x')/\tilde{W}(y|x)]^{\sigma}$, as the effect of the event
$$\left\{\tilde{W}(\by|\bx_{m^\prime})> \max_{\tilde{m}\ne
(m,m^\prime)}
\tilde{W}(\by|\bx_{\tilde{m}})\right\}$$
is negligible and hence ignored. In particular, for matched ML decoding, where
$\tilde{W}=W$, the choice $\sigma=\frac{1}{2}$ corresponds to the appearance of
the Bhattacharyya distance in this expression.
As $R$ grows, pairwise error events gradually
cease to dominate the error probability, and the decoded codeword, symbolized
by $x'$, has to ``compete'', not only with
the correct codeword, but also with all other codewords at the same time.
Indeed, the factor
$$\left[\sum_{\tilde{x}}P(\tilde{x})\tilde{W}^{1/\lambda}(y|\tilde{x})\right]^{\lambda},$$
at the denominator of (\ref{innersum}), represents the typical overall collective contribution of
all other competing codewords, except the correct one. More precisely, it
stands for the typical value of the highest likelihood score among all other
wrong codewords, which are drawn randomly and independently of the channel
output $y$. As $R$ grows beyond a certain point,
more and more weight is given to this term at the expense of the factor
pertaining to the correct codeword $x$. This means that $\tau$ ceases to be equal
to zero and it becomes positive. Another effect of increasing $R$ is via the
choice of the parameter $\lambda$.
For $\lambda\to 0$, this factor tends to
$\max_{\tilde{x}}\tilde{W}(y|\tilde{x})$, which
means that at extremely high coding rates, there is enough probability that
one of the wrong randomly generated codewords is composed of the ``most
likely'' input
letter for each coordinate of $\by$.
For $\lambda=1$, it is equal to $\sum_x P(x)\tilde{W}(y|x)$, which, for the
matched case ($\tilde{W}=W$), corresponds to
$R=I(X;Y)$, the mutual information induced by $P$ and $W$. 
Finally, in the other extreme, $\lambda\to\infty$ yields $\exp\{\sum_xP(x)\ln
\tilde{W}(y|x)\}$, which is the typical score of a single randomly chosen codeword,
namely, zero rate.

From this point onward, until the last paragraph of this section, our discussion focuses on the matched case,
namely, $\tilde{W}=W$ and $\beta\to\infty$. In this case, it is interesting to observe that
the above derived expression of the TRC exponent
is simultaneously a
generalized form of both the random coding exponent function and the form of
the expurgated function at low rates.
To see this, consider the following two
cases.\\

\noindent
1. For a given low rate $R$, let $\varrho\ge 1$ be the achiever of
\begin{equation}
\sup_{\rho\ge 1}[E_{\mbox{\tiny x}}(\rho)-2\rho R],
\end{equation}
where
\begin{equation}
E_{\mbox{\tiny x}}(\rho)=
\rho\ln\left(\sum_{x,x^\prime}P(x)P(x^\prime)
\left[\sum_y\sqrt{W(y|x)W(y|x^\prime)}\right]^{1/\rho}\right),
\end{equation}
and where by ``low rate'', we mean $R\le \frac{R_{\mbox{\tiny x}}}{2}\dfn
\frac{\dot{E}_{\mbox{\tiny
x}}(1)}{2}$, $\dot{E}_{\mbox{\tiny x}}(\rho)$ being the derivative of
$E_{\mbox{\tiny x}}(\rho)$.
Now, let $\sigma=\frac{1}{2}$, $\tau=0$
(so $\lambda$ is immaterial), $\zeta=\varrho$ and $\theta=\varrho-1$. The
matched TRC exponent is then lower bounded by
\begin{eqnarray}
E_{\mbox{\tiny trc}}^g(R,P)&\ge&
-\varrho\ln\left(\sum_{x,x^\prime}P(x)P(x^\prime)
\left[\sum_y\sqrt{W(y|x)W(y|x^\prime)}\right]^{1/\varrho}\right)
-(2\varrho-1)R\nonumber\\
&=&E_{\mbox{\tiny x}}(\varrho)-(2\varrho-1)R\nonumber\\
&=&\sup_{\rho\ge
1}\{E_{\mbox{\tiny x}}(\rho)-
(2\rho-1)R\}\nonumber\\
&=&E_{\mbox{\tiny ex}}(2R,P)+R,
\end{eqnarray}
where $E_{\mbox{\tiny ex}}(\cdot,P)$ is the expurgated exponent function
\cite{Gallager68}, \cite{VO79}. This is
in agreement with the results in \cite{BF02}. In particular, for $R=0$, $E_{\mbox{\tiny
trc}}(0,P)=E_{\mbox{\tiny ex}}(0,P)$ is achieved for $\varrho\to\infty$,
which, for the optimal $P$, is
also the optimal achievable zero--rate error exponent \cite[Sect.\ 3.7]{VO79}.

\noindent
2. For a given high rate
$R$, let now $\varrho\in(0,1]$, be the achiever of
\begin{equation}
E_{\mbox{\tiny r}}(R)=\sup_{\rho\ge 0}[E_0(\rho)-\rho R]=\sup_{0\le\rho\le
1}[E_0(\rho)-\rho R],
\end{equation}
where we recall that
\begin{eqnarray}
E_0(\rho)&=&
-\ln\left(\sum_y\left[\sum_xP(x)W^{1/(1+\rho)}(y|x)\right]^{1+\rho}\right)\nonumber\\
&=&-\ln\left[\sum_y
\exp\left\{(1+\rho)A(y,1+\rho)\right\}\right]\nonumber\\
\end{eqnarray}
with the function $A(\cdot,\cdot)$ being defined as
\begin{equation}
\label{Adef}
A(y,r)\dfn\ln\left[\sum_xP(x)W^{1/r}(y|x)\right],~~~~~r> 0,
\end{equation}
and where by ``high rate'', we mean $R\ge\dot{E}_0(1)$, $\dot{E}_0(\cdot)$
being the derivative of $E_0(\cdot)$.
This means that
$R=\dot{E}_0(\varrho)$.
Now, given $\varrho$, let us choose the parameters as follows:
$\sigma=\frac{\rho}{1+\varrho}$,
$\tau=\frac{1-\varrho}{1+\varrho}$, $\zeta=1$, and $\theta=0$.
We then have
\begin{equation}
\label{trclb}
E_{\mbox{\tiny trc}}(R,P)\ge\inf_{\lambda\ge
0}\left\{E_1(\varrho,\lambda)-\left[1-\frac{\lambda(1-\varrho)}{1+\varrho}\right]R\right\},
\end{equation}
where
\begin{eqnarray}
E_1(\varrho,\lambda)&=&-\ln\left\{\sum_y\frac{\left[\sum_xP(x)W^{1/(1+\varrho)}(y|x)\right]^2}
{\left[\sum_xP(x)W^{1/\lambda}(y|x)
\right]^{\lambda(1-\varrho)/(1+\varrho)}}\right\}\nonumber\\
&=&-\ln\left[\sum_y\exp\left\{2A(y,1+\varrho)-\frac{\lambda(1-\varrho)}{1+\varrho}
A(y,\lambda)\right\}\right].
\end{eqnarray}
To find the achiever $\lambda$ of the r.h.s.\ of eq.\ (\ref{trclb}),
we equate the derivative of the objective to
zero, i.e.,
\begin{eqnarray}
0&=&\frac{\partial
E_1(\varrho,\lambda)}{\partial\lambda}+\frac{(1-\varrho)R}{1+\varrho}\nonumber\\
&\equiv&\frac{\partial
E_1(\varrho,\lambda)}{\partial\lambda}+\frac{1-\varrho}{1+\varrho}\cdot\dot{E}_0(\varrho)\nonumber\\
&\equiv&\frac{1-\varrho}{1+\varrho}\cdot\frac{\sum_y
[A(y,\lambda)+\lambda A'(y,\lambda)]\exp\{2A(y,1+\varrho)-\lambda
(1-\varrho)A(y,\lambda)/(1+\varrho)\}}
{\sum_y \exp\{2A(y,1+\varrho)-\lambda
(1-\varrho)A(y,\lambda)/(1+\varrho)\}}-\nonumber\\
& &\frac{1-\varrho}{1+\varrho}\cdot\frac{\sum_y[A(y,1+\varrho)+
(1+\varrho)A'(y,1+\varrho)]
\exp\left\{(1+\varrho)A\left(y,1+\varrho\right)\right\}}
{\sum_y\exp\left\{(1+\varrho)A\left(y,1+\varrho\right)\right\}},
\end{eqnarray}
where $A'(y,\lambda)$ is the derivative of $A(y,\lambda)$ w.r.t. $\lambda$.
It is now easy to see that $\lambda=1+\varrho$ trivially solves this equation,
and so, under the assumption (to be discussed in the next paragraph) that this
solution provides the global
minimum of the r.h.s.\ of (\ref{trclb}), we have
\begin{eqnarray}
E_{\mbox{\tiny trc}}^g(R,P)&\ge&\inf_{\lambda\ge
0}\left\{E_1(\varrho,\lambda)-\left[1-\frac{\lambda(1-\varrho)}{1+\varrho}\right]R\right\}\nonumber\\
&=&E_1(\varrho,1+\varrho)-\left[1-\frac{(1+\varrho)(1-\varrho)}{1+\varrho}\right]R\nonumber\\
&=&E_0(\varrho)-\varrho R\nonumber\\
&=&\sup_{\rho\ge 0}[E_0(\rho)-\rho R]\nonumber\\
&=&E_{\mbox{\tiny r}}(R,P)=
E_{\mbox{\tiny sp}}(R,P).
\end{eqnarray}
Obviously, the inequality must be achieved with equality, at least for the
optimal $P$, since it coincides
with the sphere--packing upper bound to the error exponent.

Referring to the above assumption that
$\lambda=1+\varrho$ is the
minimizer of $E_1(\varrho,\lambda)$ w.r.t.\ $\lambda$, our observations are as
follows. A sufficient (though not necessary) condition for this to be the case is that
$E_1(\varrho,\lambda)$ would be convex
in $\lambda$ for fixed $\varrho$.
Consider the important special
case where $A(y,\lambda)$ is independent\footnote{This is the case, for
example, if $P$ is the uniform distribution and the columns of the matrix
$\{w_{ij}=W(j|i)\}$ are permutations of each other. Even
more specifically, this happens when $\calX=\calY$ is a group and
$W(y|x)=W(y\ominus x)$, where
$\ominus$ is the difference operation w.r.t.\ this group, namely, when $W$ is
a modulo--additive channel. More generally, this condition continues to hold as
long as all columns of $W$ are
obtained from one column by permuting only components (of that column)
pertaining to channel input symbols for which $P$ assigns the same
probability.}
of $y$ for all $\lambda$. In this case, it is easy
to prove the convexity of $E_1(\varrho,\cdot)$ as follows. Abbreviating the
notation $A(y,\lambda)$ as $A(\lambda)$, we have
\begin{eqnarray}
E_1(\varrho,\lambda)&=&-\ln\left[\sum_y\exp\left\{2A\left(1+\varrho\right)-\frac{\lambda(1-\varrho)}{1+\varrho}
A\left(\lambda\right)\right\}\right]\nonumber\\
&=&-\ln\left[|\calY|\cdot\exp\left\{2A\left(1+\varrho\right)-\frac{\lambda(1-\varrho)}{1+\varrho}
A\left(\lambda\right)\right\}\right]\nonumber\\
&=&\frac{1-\varrho}{1+\varrho}\cdot\lambda
A\left(\lambda\right)-\ln|\calY|-\ln\left[2A\left(1+\varrho\right)\right],
\end{eqnarray}
which is convex in $\lambda$ for fixed $\varrho$ because
the function $\lambda\cdot
A(\lambda)$ is such, as will be shown in Section
\ref{proof} (see, in particular, eq.\ (\ref{alphadual}) and the following text).
Beyond the class of channels with the above
described symmetry property, a numerical study indicates that
$E_1(\varrho,\cdot)$
remains convex for many other combinations of $P$ and $W$, but not in general.
For example, when $W$ is the binary z--channel, this is not always the case.

To summarize, there are basically three ranges with different kinds of behavior
of the TRC exponent. Denoting
$R_{\mbox{\tiny c1}}=\frac{\dot{E}_{\mbox{\tiny x}}(1)}{2}$ and
$R_{\mbox{\tiny c2}}=\dot{E}_0(1)$, 
we have the following explicit lower bounds to
the TRC exponent, which are coherent with the findings of \cite{BF02}, that were
derived for the special case of the binary symmetric channel and
codes drawn by fair coin tossing.
\begin{enumerate}
\item {\it Low rates.} For $R \le R_{\mbox{\tiny c1}}$, the graph of the TRC
exponent function is a convex curve, with
an initial slope of
$-\infty$ and final slope of $-1$. Here, $\sigma=\frac{1}{2}$, $\tau=0$,
and $\zeta=1+\theta$ decreases from $\infty$ to $1$ and
$$E_{\mbox{\tiny trc}}^g(R,P)\ge E_{\mbox{\tiny ex}}(2R,P)+R.$$
\item {\it Moderate rates.} For $R_{\mbox{\tiny c1}}\le R \le R_{\mbox{\tiny
c2}}$, the TRC exponent is an
affine function of $R$ with slope $-1$ (i.e., the graph is a straight line).
Here, $\sigma$ and
$\tau$ are as before, but $\theta=0$ and $\zeta=1$. In this case,
$$E_{\mbox{\tiny trc}}^g(R,P)\ge E_0(1)-R.$$
\item {\it High rates.} For $R \ge R_{\mbox{\tiny c2}}$, the graph of the TRC
exponent function is
again a convex curve, with an initial slope of $-1$ and final slope of $0$.
Here, every rate corresponds to a value of $\varrho$ that decreases from $1$
to $0$, and the parameters are: $\sigma=\frac{\varrho}{1+\varrho}$,
$\tau=\frac{1-\varrho}{1+\varrho}$, $\lambda=1+\varrho$, $\theta=0$ and
$\zeta=1$, and then
$$E_{\mbox{\tiny trc}}^g(R,P)\ge E_{\mbox{\tiny sp}}(R,P).$$
\end{enumerate}
It should be pointed out that since the parameters $\sigma$, $\tau$, $\zeta$ and
$\theta$, were chosen here in a specific (seemingly, arbitrary) manner, and not as a result of
the maximizations, the resulting expressions are merely lower bounds to the
TRC exponent, and there is no guarantee that they are the exact quantities, at
all rates, except the cases of $R = 0$ and high rates, above the critical rate, where if $P$ is chosen
optimally, these figures meet the well known zero--rate bound
and the sphere--packing bound, respectively. However, at intermediate rates,
where the exact reliability function is no fully known, 
it is not clear, for example, that the choice $\tau=0$ continues to be optimal for all rates up
to $R_{\mbox{\tiny c2}}$. We could not rule out the theoretical possibility that the 
passage from $\tau=0$ to $\tau>0$, which stands for
the point at which pairwise error events cease to dominate the error
probability, might occur
at a rate strictly lower than $R_{\mbox{\tiny c2}}$. On the other hand, numerical studies that we
have conducted so far, did not reveal examples where this in fact happens.
We therefore conjecture that the optimal value of $\tau$ is zero for all $R \le
R_{\mbox{\tiny c2}}$.

Our final remark concerns the dependence of the TRC exponent bound on $\beta$.
It is known \cite{BHKRS18}, \cite{LCV17} that the error probability of the matched likelihood decoder
($\tW=W$, $\beta=1$) cannot be larger than twice the error probability of the
ML decoder, and therefore, in this case, the optimal error exponent is achieved
for every $\beta\ge 1$. It seems to be less obvious, however, that 
the optimal error exponent of the ML decoder is achieved even if one starts sweeping $\beta$
from values less than 1.
Indeed, as mentioned earlier, when $\beta\to\infty$, the maximization over both
$\sigma$ and $\tau$ take place on the entire positive real line. Let
$\sigma^*$ and $\tau^*$ denote the maximizers for $\beta\to\infty$. If
$\sigma^*$ and $\tau^*$ are both finite, these
maximizers will be achieved as well whenever $\beta\ge\beta_0=\sigma^*+\tau^*$. Thus,
for low rates, if $\sigma^*=\frac{1}{2}$ and $\tau^*=0$, then
$\beta_0=\frac{1}{2}$ is the critical value of $\beta$ beyond which the error
exponent ceases to improve and remains fixed. For high rates,
if $\sigma^*=\frac{\varrho}{1+\varrho}$ and $\tau^*=\frac{1-\varrho}{1+\varrho}$,
then $\beta_0=\frac{1}{1+\varrho}$, so here $\beta_0\to 1$ only when $R\to
I(X;Y)$. Even less obvious is the similar behavior for the mismatched likelihood decoder. The first
non--trivial fact about stochastic mismatched decoding 
is that the error exponent must be a monotonically non--decreasing
function of $\beta$. The second point is that
here too, for the same reasons, if the achievers $\sigma^*$ and $\tau^*$ are
finite, the resulting error exponent would cease to depend on $\beta$ for all
$\beta\ge \beta_0=\sigma^*+\tau^*$. However, here we do not claim that
$\beta_0\le 1$ in general.

\section{Proof}
\label{proof}

Before passing to the Lagrange--dual, we first need to modify the
Csisz\'ar--style expression
of the TRC exponent, in Theorem 1 above, in order to account for the fact that we are
replacing the fixed--composition ensemble of \cite{trc} to the i.i.d.\
ensemble under $P$, as described in Section \ref{nab}.
There are only a few modifications. The first is that the constraints
$Q_{X^\prime}=Q_X$ (in eq.\ (\ref{alpha-def})) and $Q_{\tilde{X}}=Q_X$ (in
(\ref{trcee})) are now removed since the types of
the randomly chosen codewords may fluctuate around $P$. The second
modification is that the mutual information term,
$I_Q(X;X^\prime)$, in the objective of (\ref{trcee}), and
$I_Q(\tilde{X};Y)$, in both the constraint and the objective of
(\ref{alpha-def}), are replaced by $J_Q(X;X^\prime)+D(Q_X\|P)$ and $J_Q(\tilde{X};Y)$, respectively,
where
\begin{equation}
J_Q(X;X^\prime)=I_Q(X;X^\prime)+D(Q_{X^\prime}\|P)=\bE_Q\ln\frac{Q_{X^\prime|X}(X^\prime|X)}{P(X^\prime)},
\end{equation}
and
\begin{equation}
J_Q(\tilde{X};Y)=I_Q(\tilde{X};Y)+D(Q_{\tilde{X}}\|P)=\bE_Q\ln\frac{Q_{\tilde{X}|Y}(X|Y)}{P(\tilde{X})}.
\end{equation}
As for the mutual information term in the constraint of (\ref{trcee}), the
situation is slightly more involved. Referring to the proof of \cite[Theorem
1]{trc}, we have to analyze once again the moments of the type class
enumerators,
\begin{equation}
N(Q_{XX^\prime})=\sum_{m=0}^{M-1}\sum_{m^\prime\ne
m}\calI\{(\bx_m,\bx_{m^\prime})\in\calT(Q_{XX^\prime})\}.
\end{equation}
Following \cite{trc}, let us also denote by $N(Q_{XX^\prime}|\bx_m)$
the number of codewords $\{\bx_{m^\prime},~m^\prime\ne m\}$ such that 
$(\bx_m,\bx_{m^\prime})\in\calT(Q_{XX^\prime})$ (i.e., the same definition as
$N(Q_{XX^\prime})$ but without the summation over $m$). 
Similarly as in \cite[eq.\ (36)]{trc}, for given $\rho\ge s\ge 1$, 
\begin{eqnarray}
\bE\{N^{1/\rho}(Q_{XX^\prime})\}&=&
\bE\left\{\left[\sum_{m=0}^{M-1}N(Q_{XX^\prime}|\bX_m)\cdot
\calI\{\bX_m\in\calT(Q_X)\}\right]^{1/\rho}\right\}\nonumber\\
&=&\bE\left\{\left(\left[\sum_{m=0}^{M-1}N(Q_{XX^\prime}|\bX_m)\cdot
\calI\{\bX_m\in\calT(Q_X)\}\right]^{1/s}\right)^{s/\rho}\right\}\nonumber\\
&\le&\bE\left\{\left(\sum_{m=0}^{M-1}N^{1/s}(Q_{XX^\prime}|\bX_m)\cdot
\calI\{\bX_m\in\calT(Q_X)\}\right)^{s/\rho}\right\}\nonumber\\
&\le&\left(\bE\sum_{m=0}^{M-1}N^{1/s}(Q_{XX^\prime}|\bX_m)\cdot
\calI\{\bX_m\in\calT(Q_X)\}\right)^{s/\rho}\nonumber\\
&=&e^{nRs/\rho}\left(\bE\left\{N^{1/s}(Q_{XX^\prime}|\bX_m)\cdot
\calI\{\bX_m\in\calT(Q_X)\}\right\}\right)^{s/\rho}\nonumber\\
&=&e^{nRs/\rho}\left(\bE\left\{N^{1/s}(Q_{XX^\prime}|\bX_m)\right\}\cdot
P[\calT(Q_X)]\right)^{s/\rho}\nonumber\\
&\exe&e^{n[R-D(Q_X\|P)]s/\rho}\left(\bE\left\{N^{1/s}(Q_{XX^\prime}|\bX_m)\right\}
\right)^{s/\rho}\nonumber\\
&\exe&e^{n[R-D(Q_X\|P)]s/\rho}\cdot\left\{\begin{array}{ll}
e^{n[R-J_Q(X;X^\prime)]} & R> J_Q(X;X^\prime)\\
e^{n[R-J_Q(X;X^\prime)]s/\rho} & R<
J_Q(X;X^\prime)\end{array}\right.\nonumber\\
&=&\left\{\begin{array}{ll}
e^{n\{[R-D(Q_X\|P)]s/\rho+[R-J_Q(X;X^\prime)]/\rho\}} & R> J_Q(X;X^\prime)\\
e^{n[2R-D(Q_X\|P)-J_Q(X;X^\prime)]s/\rho} & R<
J_Q(X;X^\prime)\end{array}\right.\nonumber\\
\end{eqnarray}
and after minimization over $s\in[1,\rho]$:
\begin{equation}
\bE\{N^{1/\rho}(Q_{XX^\prime})\}\lexe
\left\{\begin{array}{ll}
e^{n[2R-D(Q_X\|P)-J_Q(X;X^\prime)]/\rho} & R> J_Q(X;X^\prime),~R>D(Q_X\|P)\\
e^{n[R-D(Q_X\|P)+(R-J_Q(X;X^\prime)/\rho]} & R> J_Q(X;X^\prime),~R<D(Q_X\|P)\\
e^{n[2R-D(Q_X\|P)-J_Q(X;X^\prime)]/\rho} & R<
J_Q(X;X^\prime),~2R>D(Q_X\|P)+J_Q(X;X^\prime)\\
e^{n[2R-D(Q_X\|P)-J_Q(X;X^\prime)]} & R<
J_Q(X;X^\prime),~2R<D(Q_X\|P)+J_Q(X;X^\prime)\end{array}\right.\nonumber
\end{equation}
Now,
\begin{eqnarray}
& &\lim_{\rho\to\infty}\left[\bE\{N^{1/\rho}(Q_{XX^\prime})\}\right]^\rho\nonumber\\&\le&
\left\{\begin{array}{ll}
e^{n[2R-D(Q_X\|P)-J_Q(X;X^\prime)]} & R> J_Q(X;X^\prime),~R>D(Q_X\|P)\\
0 & R> J_Q(X;X^\prime),~R<D(Q_X\|P)\\
e^{n[2R-D(Q_X\|P)-J_Q(X;X^\prime)]} & R<
J_Q(X;X^\prime),~2R>D(Q_X\|P)+J_Q(X;X^\prime)\\
0 & R<
J_Q(X;X^\prime),~2R<D(Q_X\|P)+J_Q(X;X^\prime)\end{array}\right.\nonumber\\
&=&\left\{\begin{array}{ll}
e^{n[2R-D(Q_X\|P)-J_Q(X;X^\prime)]} & 2R > F_Q(X,X^\prime)\\
0 & 2R < F_Q(X,X^\prime)\end{array}\right.
\end{eqnarray}
where
\begin{equation}
F_Q(X,X^\prime)=D(Q_X\|P)+\max\{D(Q_X\|P),J_Q(X;X^\prime)\}.
\end{equation}

Using these facts the same way as in \cite{trc}, we find that
the TRC exponent for the i.i.d.\ ensemble and
$g(Q)=\beta\bE_Q\ln\tW(Y|X)$, is as follows.
We first ref--define $\alpha(R,Q_Y)$ as
\begin{equation}
\label{alpha-def-modified}
\alpha(R,Q_Y)\dfn\sup_{\{Q_{\tilde{X}|Y}:~J_Q(\tilde{X};Y)\le R
\}}[\beta\bE_Q\ln\tW(Y|\tX)-J_Q(\tilde{X};Y)]+R,
\end{equation}
$\Gamma(Q_{XX^\prime},R)$ is defined as in (\ref{Gamma-def}) (but with
$g(Q)=\beta\bE_Q\ln\tW(Y|X)$), and
finally,
\begin{equation}
\label{trcee-modified}
E_{\mbox{\tiny trc}}^g(R,P)\ge\inf_{\{Q_{X^\prime|X}:~F_Q(X;X^\prime)\le
2R\}}\{\Gamma(Q_{XX^\prime},R)+
J_Q(X;X^\prime)+D(Q_X\|P)-R\}.
\end{equation}
We are now ready to move on to the main part of the proof, which is the
derivation of the Lagrange--dual form of this lower bound to $E_{\mbox{\tiny trc}}^g(R,P)$.

Throughout this proof we will make a frequent use of the minimax theorem, based
on convexity--concavity arguments. We will also use repeatedly the fact that
for a given function $f:\calX\to\reals$ that does not depend on $Q$,
$$\min_Q[D(Q\|P)+\bE_Q\{f(X)\}]=-\ln\bE_P\left\{e^{-f(X)}\right\},$$
which can easily be verified either by carrying out the minimization
using standard methods, or by writing the the objective on the l.h.s.\ as
$$D(Q\|P')-\ln\bE_P\left\{e^{-f(X)}\right\},$$
with $P'(x)\propto P(x)e^{-f(x)}$, which is obviously minimized by $Q=P'$.

We begin from $\alpha(R,Q_Y)$, which is the inner--most optimization.
\begin{eqnarray}
\label{alphadual}
\alpha(R,Q_Y)&=&\sup_{\{Q_{\tX|Y}:~J_Q(\tX;Y)\le
R\}}[\beta\bE_Q\ln\tW(Y|\tX)-J_Q(\tX;Y)]+R\nonumber\\
&=&\sup_{Q_{\tX|Y}}\inf_{\lambda > 0}\left[\sum_yQ_Y(y)\sum_x Q_{\tX|Y}(x|y)\ln
\tW^\beta(y|x)+\right.\nonumber\\
& &\left.\lambda\left(R-\sum_x
Q_{\tX|Y}(x|y)\ln\frac{Q_{\tX|Y}(x|y)}{P(x)}\right)\right]\nonumber\\
&=&\inf_{\lambda > 0}\sup_{Q_{\tX|Y}}\lambda\cdot\left[\sum_yQ_Y(y)\sum_x
Q_{\tX|Y}(x|y)\ln
\frac{P(x)\tW^{\beta/\lambda}(y|x)}{Q_{\tX|Y}(x|y)}+R
\right]\nonumber\\
&=&\inf_{\lambda > 0}\lambda\cdot\left[\sum_yQ_Y(y)\ln\left(\sum_x
P(x)\tW^{\beta/\lambda}(y|x)\right)+R\right]\nonumber\\
&=&\inf_{\lambda >
0}\lambda\cdot\left[\sum_yQ_Y(y)A\left(y,\frac{\lambda}{\beta}\right)+R\right],
\end{eqnarray}
where the function $A$ has been defined in (\ref{Adef}).
Observe that $\lambda\cdot A(y,\lambda/\beta)$ is always a
convex function, since it
was obtained as the supremum (over $\{Q_{\tX|Y}\}$) of affine functions in
$\lambda$. Now,
\begin{eqnarray}
E_{\mbox{\tiny trc}}^g(R,P)&\ge&\inf_{\{Q_{XX^\prime Y}:~F_Q(X,X^\prime)\le
2R\}}\{D(Q_{Y|X}\|W|Q_X)+\nonumber\\
& &I_Q(X^\prime;Y|X)+J_Q(X;X^\prime)+D(Q_X\|P)+\nonumber\\
& &[\max\{\beta\bE_Q\ln\tW(Y|X),\alpha(R,Q_Y)\}-\beta\bE_Q\ln\tW(Y|X^\prime)]_+\}-R\nonumber\\
&=&\inf_{\{Q_{XX^\prime Y}:~F_Q(X,X^\prime)\le
2R\}}\bigg\{-\bE_Q\ln W(Y|X)-
H_Q(Y|X,X^\prime)+\nonumber\\
& &J_Q(X;X^\prime)+D(Q_X\|P)+\nonumber\\
& &[\max\{\beta\bE_Q\ln\tW(Y|X),\alpha(R,Q_Y)\}-\beta\bE_Q\ln\tW(Y|X^\prime)]_+
\bigg\}-R\nonumber\\
&=&\inf_{\{Q_{XX^\prime Y}:~F_Q(X,X^\prime)\le
2R\}}\sup_{0\le s\le
1}\sup_{0\le t \le 1}\inf_{\lambda > 0}\bigg\{J_Q(X;X')+D(Q_X\|P)+\nonumber\\
& &\sum_{x,x^\prime}Q_{XX^\prime}(x,x^\prime)
\sum_yQ_{Y|XX^\prime}(y|x,x^\prime)\left[
\ln\frac{Q_{Y|XX^\prime}(y|xx^\prime)}{W(y|x)}+\right.\nonumber\\
& &\left.s\left(t\ln
\tW^\beta(y|x)+(1-t)\lambda[A(y,\lambda/\beta)+R]-\ln\tW^\beta(y|x')\right)\right]\bigg\}-R\nonumber\\
&=&\inf_{\{Q_{XX^\prime Y}:~F_Q(X,X^\prime)\le
2R\}}\sup_{0\le s\le
1}\sup_{0\le r \le s}\inf_{\lambda > 0}\bigg[J_Q(X;X')+D(Q_X\|P)+\nonumber\\
& &\sum_{x,x^\prime}Q_{XX^\prime}(x,x^\prime)\sum_yQ_{Y|XX^\prime}(y|xx^\prime)\left(
\ln\frac{Q_{Y|XX^\prime}(y|xx^\prime)}{W(y|x)}+\right.\nonumber\\
& &\left.r\ln\tW^\beta(y|x)+
(s-r)\lambda[A(y,\lambda/\beta)+R]-s\ln\tW^\beta(y|x')\right)\bigg]-R\nonumber\\
&\ge&\sup_{0\le s\le
1}\sup_{0\le r \le s}\inf_{\lambda >
0}\inf_{\{Q_{XX^\prime Y}:~F_Q(X,X^\prime)\le
2R\}}\bigg[J_Q(X;X')+D(Q_X\|P)+\nonumber\\
& &\sum_{x,x^\prime}Q_{XX^\prime}(x,x^\prime)\sum_yQ_{Y|XX^\prime}(y|xx^\prime)\times\nonumber\\
& &\left(\ln\frac{Q_{Y|XX^\prime}(y|xx^\prime)\tW^{\beta
r}(y|x)}{W(y|x)\tW^{\beta s}(y|x^\prime)e^{-\lambda(s-r)A(y,\lambda/\beta)}}
-[1-\lambda(s-r)]R\right)\bigg]\nonumber\\
&=&\sup_{0\le s\le
1}\sup_{0\le r \le s}\inf_{\lambda >
0}\inf_{\{Q_{XX^\prime}:~F_Q(X,X^\prime)\le
2R\}}\bigg\{J_Q(X;X')+D(Q_X\|P)+\nonumber\\
& &\sum_{x,x^\prime}Q_{XX^\prime}(x,x^\prime)\left(-\ln\left[
\sum_yW(y|x)\cdot\frac{\tW^{\beta s}(y|x^\prime)}{
\tW^{\beta r}(y|x)e^{\lambda(s-r)A(y,\lambda/\beta)}}\right]
-[1-\lambda(s-r)]R\right)\bigg\}\nonumber\\
&=&\sup_{0\le s\le
1}\sup_{0\le r \le s}\inf_{\lambda >
0}\inf_{\{Q_{XX^\prime}:~F_Q(X,X^\prime)\le
2R\}}\bigg[J_Q(X;X')+D(Q_X\|P)+
\sum_{x,x^\prime}Q_{XX^\prime}(x,x^\prime)\times\nonumber\\
& &\left\{-\ln\left(
\sum_yW(y|x)\cdot
\frac{\tW^{\beta s}(y|x^\prime)}{\tW^{\beta r}(y|x)\left[\sum_{\tilde{x}}P(\tilde{x})
\tW^{\beta/\lambda}(y|\tilde{x})\right]^{\lambda(s-r)}}\right)
-[1-\lambda(s-r)]R\right\}.\nonumber
\end{eqnarray}
Let us denote
\begin{equation}
G(x,x^\prime,\lambda,s,r)\dfn-\ln\left(
\sum_yW(y|x)\cdot\frac{\tW^{\beta s}(y|x^\prime)}{\tW^{\beta r}(y|x)\left[\sum_{\tilde{x}}P(\tilde{x})
W^{1/\lambda}(y|\tilde{x})\right]^{\lambda(s-r)}}\right).
\end{equation}
Then,
\begin{eqnarray}
E_{\mbox{\tiny trc}}^g(R,P)&\ge&\sup_{0\le s\le 1}
\sup_{0\le r \le s}\inf_{\lambda \ge 0}
\inf_{\{Q_{XX^\prime}:~F_Q(X,X^\prime)\le
2R\}}\bigg[J_Q(X;X')+D(Q_X\|P)+\nonumber\\
& &\sum_{x,x^\prime}Q_{XX^\prime}(x,x^\prime)\left\{G(x,x^\prime,\lambda,s,r)-[1-\lambda(s-r)]
R\right\}\nonumber\\
&=&\sup_{0\le s\le 1}
\sup_{0\le r \le s}\inf_{\lambda \ge
0}\inf_{\{Q_{XX^\prime}:~F_Q(X,X^\prime)\le
2R\}}\sum_{x,x^\prime}Q_{XX^\prime}(x,x^\prime)\left(
\ln\frac{Q_{XX^\prime}(x,x^\prime)}{P(x)P(x^\prime)}+\right.\nonumber\\
& &\left.G(x,x^\prime,\lambda,s,r)-
[1-\lambda(s-r)]R\right)\nonumber\\
&=&\sup_{0\le s\le 1}
\sup_{0\le r \le s}\inf_{\lambda \ge 
0}\inf_{Q_{XX^\prime}}\sup_{\zeta \ge 0}\sup_{\theta\ge 0}
\sum_{x,x^\prime}Q_{XX^\prime}(x,x^\prime)\left[
\ln\frac{Q_{XX^\prime}(x,x^\prime)}{P(x)P(x^\prime)}+\right.\nonumber\\
& &\left.\zeta\left(\ln\frac{Q_X(x)}{P(x)}-R\right)+\theta\left(
\ln\frac{Q_{XX^\prime}(x,x^\prime)}{P(x)P(x^\prime)}-2R\right)+\right.\nonumber\\
& &\left.G(x,x^\prime,\lambda,s,r)-\{1-\lambda(s-r)\}R\right]\nonumber\\
&=&\sup_{0\le s\le 1}
\sup_{0\le r \le s}\inf_{\lambda\ge
0}\sup_{\zeta \ge 0}\sup_{\theta\ge 0}\inf_{Q_{XX^\prime}}
\sum_{x,x^\prime}Q_{XX^\prime}(x,x^\prime)\left[
\ln\frac{Q_{XX^\prime}(x,x^\prime)}{P(x)P(x^\prime)}+\right.\nonumber\\
& &\left.\zeta\left(\ln\frac{Q_X(x)}{P(x)}-R\right)+\theta\left(
\ln\frac{Q_{XX^\prime}(x,x^\prime)}{P(x)P(x^\prime)}-2R\right)+\right.\nonumber\\
& &\left.G(x,x^\prime,\lambda,s,r)-\{1-\lambda(s-r)\}R\right]\nonumber\\
&=&\sup_{0\le s\le 1}
\sup_{0\le r \le s}\inf_{\lambda \ge
0}\sup_{\zeta \ge 0}\sup_{\theta\ge 0}\inf_{Q_{XX^\prime}}
\sum_{x,x^\prime}Q_{XX^\prime}(x,x^\prime)\left[
(1+\theta)\ln\frac{Q_{XX^\prime}(x,x^\prime)}{P(x)P(x^\prime)}+\right.\nonumber\\
& &\left.\zeta\ln\frac{Q_X(x)}{P(x)}+
G(x,x^\prime,\lambda,s,r)-\{1-\lambda(s-r)+2\theta+\zeta\}R\right]\nonumber\\
&=&\sup_{0\le s\le 1}
\sup_{0\le r \le s}\inf_{\lambda \ge
0}\sup_{\zeta \ge 0}\sup_{\theta\ge
0}\inf_{Q_X}\sum_{x}Q_X(x)\left[(1+\theta+\zeta)\ln\frac{Q_X(x)}{P(x)}+
(1+\theta)\inf_{Q_{X^\prime|X}}\right.\nonumber\\
& &\left.\left(\sum_{x^\prime}Q_{X^\prime|X}(x^\prime|x)
\ln\frac{Q_{X^\prime|X}(x^\prime|x)}{P(x^\prime)}+
G(x,x^\prime,\lambda,s,r)-\{1-\lambda(s-r)+2\theta+\zeta\}R\right)\right]\nonumber\\
&=&\sup_{0\le s\le 1}
\sup_{0\le r \le s}\inf_{\lambda \ge
0}\sup_{\zeta \ge 0}\sup_{\thetaḱ\ge
0}\inf_{Q_X}\sum_{x}Q_X(x)\left[(1+\theta+\zeta)\ln\frac{Q_X(x)}{P(x)}-\right.\nonumber\\
& &\left.(1+\theta)\ln\left(
\sum_{x^\prime}P(x^\prime)e^{-G(x,x^\prime,\lambda,s,r)/(1+\theta)}\right)-
\{1-\lambda(s-r)+2\theta+\zeta\}R\right].
\end{eqnarray}
Denoting
\begin{equation}
T(x,\lambda,s,r,\theta)\dfn\ln\left(\sum_{x^\prime}P(x^\prime)
e^{-G(x,x^\prime,\lambda,s,r)/(1+\theta)}\right),
\end{equation}
we finally, have
\begin{eqnarray}
E_{\mbox{\tiny trc}}^g(R,P)&\ge&
\sup_{0\le s\le 1}
\sup_{0\le r \le s}\inf_{\lambda\ge
0}\sup_{\zeta \ge 0}\sup_{\theta\ge
0}\bigg\{-(1+\theta+\zeta)\ln\bigg[\sum_xP(x)e^{(1+\theta)T(x,\lambda,s,r,\theta)/(1+\theta+\zeta)}
\bigg]-\nonumber\\
& &[1-\lambda(s-r)+2\theta+\zeta]R\bigg\}\nonumber\\
&=&\sup_{0\le s\le 1}
\sup_{0\le r \le s}\inf_{\lambda\ge
0}\sup_{\zeta\ge 0}\sup_{\theta\ge
0}\left[-(1+\theta+\zeta)\ln\left\{\sum_xP(x)\left[\sum_{x^\prime}
P(x^\prime)\times\right.\right.\right.\nonumber\\
& &\left.\left.\left.\left(\sum_yW(y|x)\cdot\frac{\tW^{\beta
s}(y|x^\prime)}{\tW^{\beta r}(y|x)\left[\sum_{\tilde{x}}P(\tilde{x})
tW^{\beta/\lambda}(y|\tilde{x})\right]^{\lambda(s-r)}}\right)^{1/(1+\theta)}\right]^{(1+\theta)/(1+\theta
+\zeta)}\right\}-\right.\nonumber\\
& &\left.[1-\lambda(s-r)+2\theta+\zeta]R\right]\nonumber\\
&=&\sup_{0\le \sigma \le \beta}
\sup_{0\le \tau\le \beta-\sigma}\inf_{\lambda \ge
0}\sup_{\theta\ge
0}\sup_{\zeta \ge 1+\theta}\left[-\zeta\ln\left\{\sum_xP(x)\left[\sum_{x^\prime}
P(x^\prime)\times\right.\right.\right.\nonumber\\
& &\left.\left.\left.\left(\sum_yW(y|x)\frac{\tW^{\sigma+\tau}(y|x^\prime)}{\tW^{\sigma}
(y|x)\left[\sum_{\tilde{x}}P(\tilde{x})
\tW^{1/\lambda}(y|\tilde{x})\right]^{\lambda
\tau}}\right)^{1/(1+\theta)}\right]^{(1+\theta)/\zeta}
\right\}-\right.\nonumber\\
& &\left.(\zeta+\theta-\lambda\tau)R\right],
\end{eqnarray}
where in the last step, 
we have changed 
parameters according $\beta r\to \sigma$, $\beta s\to \sigma+\tau$,
$\beta(s-r)=\tau$, and we have re--defined 
$\lambda/\beta$ as $\lambda$, and $1+\theta+\zeta$ as $\zeta$.
This completes the proof of Theorem 1.

%\section*{Appendix}
%\renewcommand{\theequation}{A.\arabic{equation}}
%    \setcounter{equation}{0}

\end{document}